\documentclass[utf8]{frontiersFPHY} 
\setcitestyle{square} 
\usepackage{url,hyperref,lineno,microtype,subcaption}
\usepackage[onehalfspacing]{setspace}
\usepackage{amsmath}
\usepackage{booktabs}
\usepackage{multirow}
\usepackage{braket}
\usepackage{caption}
\usepackage{color}
\usepackage{euscript} 

\newcommand\CO{{\mathcal O}}

\newcommand{\be}{\begin{equation}}
\newcommand{\ee}{\end{equation}}

\def\keyFont{\fontsize{8}{11}\helveticabold }
\def\firstAuthorLast{Acharya {et~al.}} 
\def\Authors{Bijaya Acharya$^{1,2}$, Sonia Bacca\,$^{1,3}*$, Francesca Bonaiti$^1$, Simone Salvatore Li Muli\,$^{1}$,
Joanna E. Sobczyk$^{1}$}
\def\Address{$^{1}$Institut f\"ur Kernphysik and PRISMA$^+$ Cluster of Excellence, Johannes Gutenberg Universit\"at, 55128
  Mainz, Germany
  
  $^{2}$ Physics Division, Oak Ridge National Laboratory, Oak
  Ridge, Tennessee 37831, USA
  
$^{3}$Helmholtz-Institut Mainz, Johannes Gutenberg Universit\"at Mainz, D-55099 Mainz, Germany
  }
\def\corrAuthor{Sonia Bacca}

\def\corrEmail{s.bacca@uni-mainz.de}

\begin{document}
\onecolumn
\firstpage{1}

\title{Uncertainty quantification in electromagnetic observables of nuclei} 

\author[\firstAuthorLast ]{\Authors} 
\address{} 
\correspondance{} 

\extraAuth{}


\maketitle

\begin{abstract}
We present strategies to quantify theoretical uncertainties in modern ab-initio calculations of electromagnetic observables in light and medium-mass nuclei. We discuss how uncertainties build up from various sources, such as the approximations introduced by the few- or many-body solver and  the truncation of the chiral effective field theory expansion.
We review the recent progress encompassing a broad range of electromagnetic observables in stable and unstable nuclei.

\tiny
 \keyFont{ \section{Keywords:} uncertainty quantification, electromagnetic processes,  ab-initio theory,  chiral effective field theory}
\end{abstract}

\section{Introduction}

Uncertainty quantification is an emerging field in nuclear theory. It is nowadays expected for any theoretical calculation of nuclear observables to have a corresponding uncertainty bar,  which is vital to make progress in our understanding of strongly interacting systems through the comparison of theoretical modeling with experimental data. While this is clearly the goal, the specific approach to uncertainty quantification and its sophistication level strongly depends on the used theoretical method and on the observables under investigation.
In this review, we focus on electromagnetic reactions and on how they can be calculated with corresponding uncertainty in the so-called ab-initio methods. It is fair to say that the sub-field of quantification of theoretical uncertainties is just now developing, and while there is still much to be done there has been recent significant progress. Here, we report on such progress, discuss its philosophy and identify areas where improvements can be expected in the future.

In the ab-initio approach to nuclear theory~\cite{Leidemann2012,BaccaPastore2014,Hebeler2015} the goal is to explain nuclear phenomena, including  electromagnetic processes, starting from protons and neutrons as degrees of freedom and to solve the related quantum-mechanical problem 
in a numerical way, either exactly or within controlled approximations. To achieve this, one typically solves the Schr\"odinger equation for a given Hamiltonian $H$  and then computes transition matrix elements of the electromagnetic operator $J^\mu$ between the eigenstates of $H$. Hence, before discussing the approach devised to quantify uncertainty in electromagnetic observables, we define the  dynamical ingredients (Hamiltonian and currents), as well as the specific observables we want to investigate.

\vspace{1cm}

\subsection{Hamiltonians and currents}
\begin{figure}[hbt]
\centering
	\includegraphics[width=0.7\textwidth]{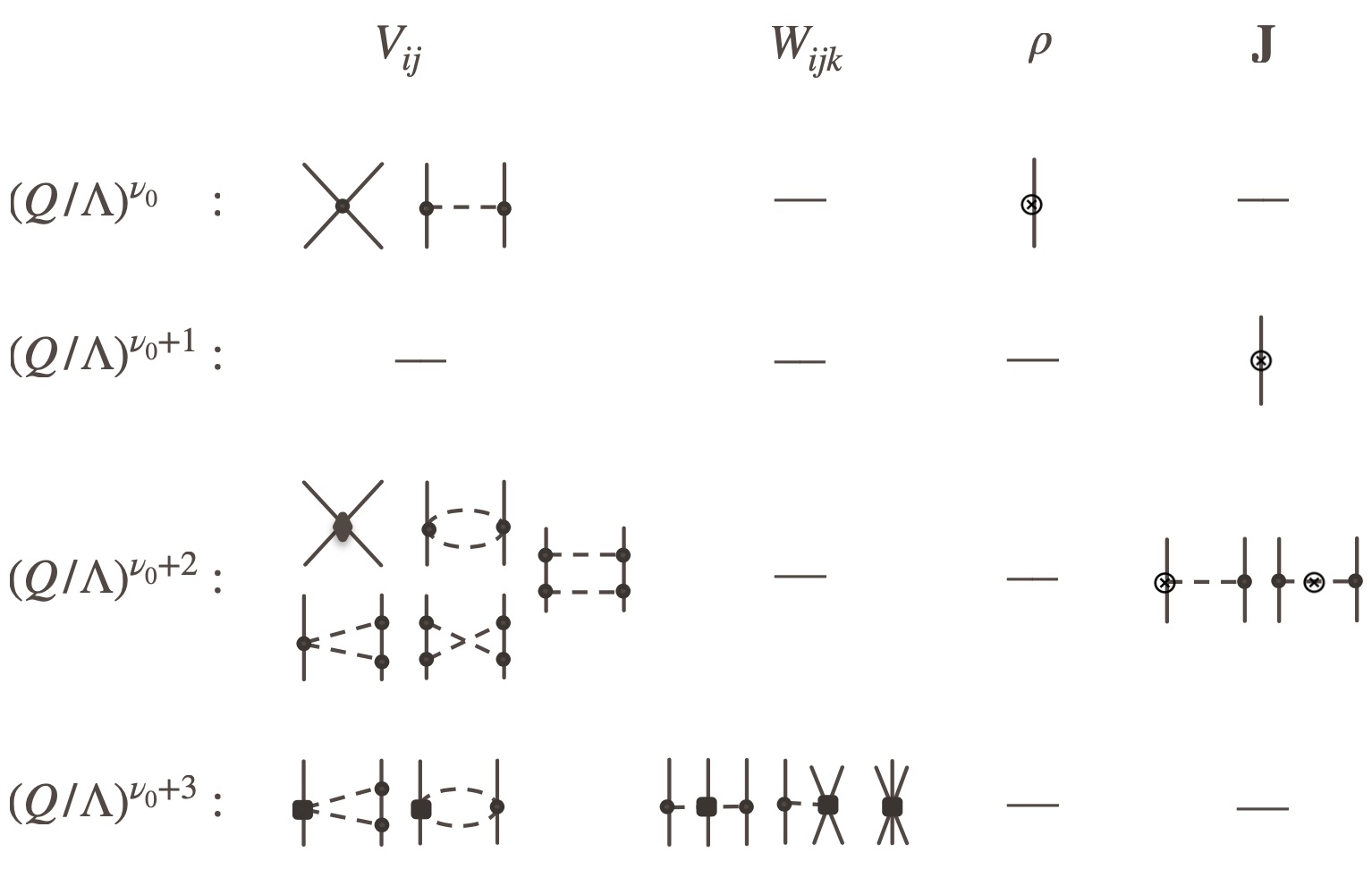}
	\caption{The $\chi$EFT expansion of the nuclear Hamiltonian and electromagnetic currents. The filled circles, squares and diamond denote strong-interaction vertices with chiral dimension $0,1$ and 2, respectively. The $\otimes$ symbols denote the electromagnetic vertices. In the literature, $\nu_0$ is usually taken as $0$ for the potential and $-3$ for the currents.}
	\label{fig_H_J}
\end{figure}

The starting point of an ab-initio computation of a nucleus composed of $A$ nucleons
is the nuclear Hamiltonian, 
\begin{equation}
\label{H}
H = T_K + \sum _{i<j}^A V_{ij} + \sum_{i<j<k}^A W_{ijk}  \,,  
\end{equation}
where $T_K$ is the intrinsic kinetic energy,
$V_{ij}$ is the two-body interaction and $W_{ijk}$ is the three-body interaction. As opposed to a phenomenological derivation of nuclear forces,  effective field theories (EFT) offer a more systematic approach~\cite{hammer2000}. In this paper, we will use effective Hamiltonians which are derived in chiral effective field theory ($\chi$EFT)\cite{Weinberg90,Epelbaum09,Machleidt11}. In this framework, the Hamiltonian is expanded
in powers of $(Q/\Lambda)$, where $Q$ is the typical low--momentum characterizing  
nuclear physics  and  $\Lambda$ is the breakdown  scale of the effective field theory. The various components relevant for $V_{ij}$ and $ W_{ijk}$ are presented in terms of Feynman diagrams in Figure~\ref{fig_H_J}, where $\nu_0$ is the first power entering in the counting.
The unresolved short range physics is encoded in the values of the  low energy constants (LECs), which are usually calibrated by fitting to experimental data. Different optimization and fitting strategies have been used to calibrate the LECs~\cite{ekstrom2013,ekstrom2015a,navarro-perez,piarulli_deltafull}. Here, we will use only a selected set of different Hamiltonians obtained from $\chi$EFT. Furthermore, interactions with explicit $\Delta$ degrees of freedom are  becoming available~\cite{kaiser1998,krebs2007,epelbaum2008,piarulli2016,ekstrom2017,jiang2020} and should be explored.
In the present work we will present results with  both chiral $\Delta$-full and $\Delta$-less interactions.

The nuclear response to external probes is described by the interaction Hamiltonian, which depends on nuclear dynamics through the nuclear current operator. The $\chi$EFT expansion exists also for the electromagnetic four-vector current $J^{\mu}=(\rho, {\bf J})$, where the time-like component is the charge operator and the space-like component is the three-vector current operator. The first diagram entering the $\chi$EFT expansion for $(\rho, {\bf J})$ are shown in Figure~\ref{fig_H_J}, where we omit the diagrams that contribute to the elastic form factors. The reader can find more details on our implementation of the currents in Ref.~\cite{bijaya}. 
While different authors adopt different power counting schemes for the currents~\cite{Pastore:2008ui,pastore2009,koelling2009,phillips2016_review}, we follow the conventions of Ref.~\cite{phillips2016_review}. 

\vspace{0.5cm}
\subsection{Electromagnetic observables}
\begin{figure}[hbt]
\centering
	\includegraphics[width=0.5\textwidth]{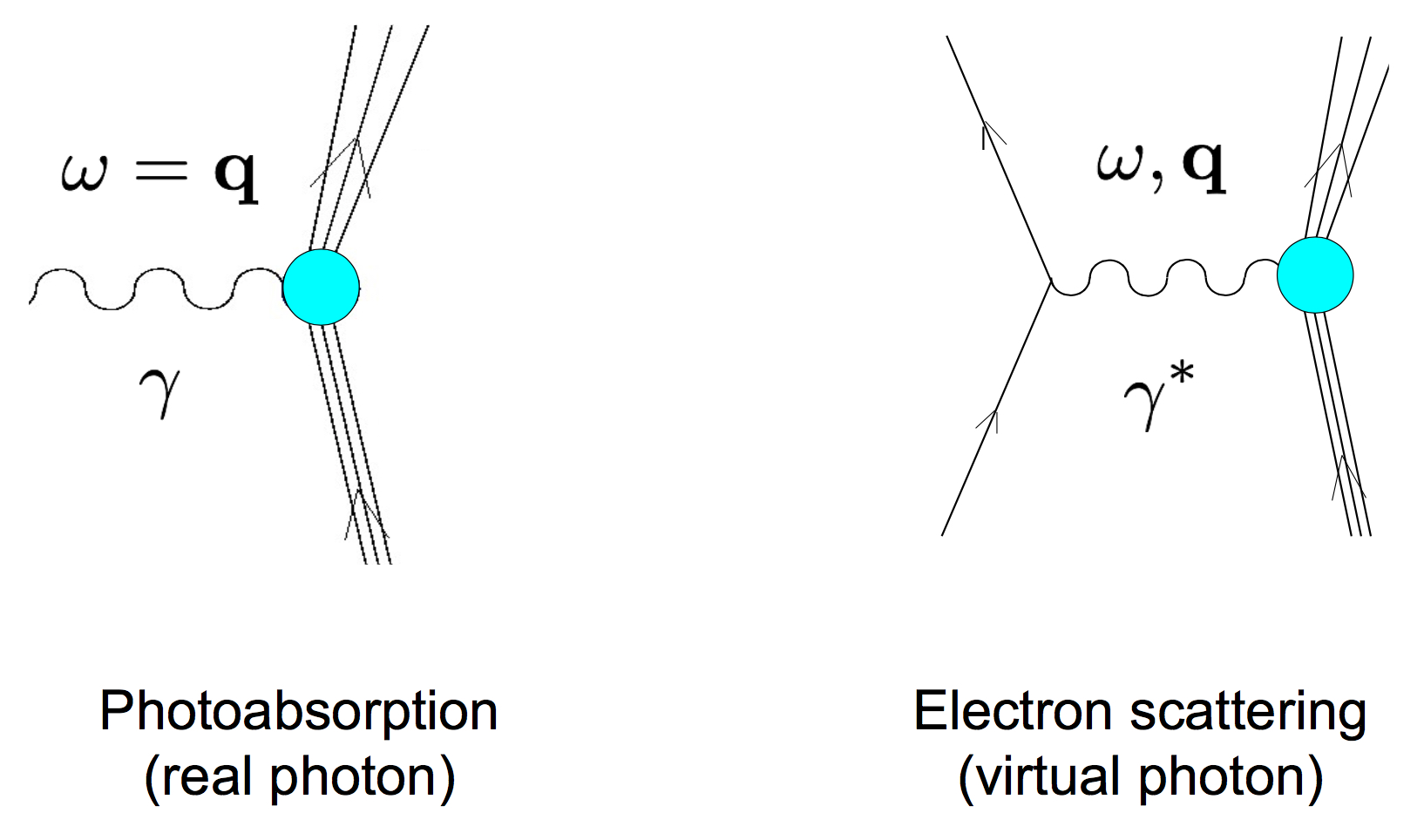}
	\caption{Feynman diagrams for the photoabsorption process (left), where a real photon $\gamma$ is exchanged, and the electron scattering process, where a virtual photon $\gamma^*$ is exchanged between the probe and the nucleus (cyan blob).}
	\label{fig_em}
\end{figure}

Electromagnetic probes are key tools to study nuclear structure because
measured cross sections are easily related to the few-/many-body matrix elements of electromagnetic operators via perturbation theory. Here, we focus on electromagnetic observables that can be explained to high precision in first order perturbation theory, i.e., processes where one single photon is exchanged between the probe and the nucleus. This is the case for the photoabsorption process and the electron scattering process, see Figure~\ref{fig_em}.
 The exchanged photon can in general transfer energy $\omega$ and momentum ${\bf q}$. In the photonuclear process, a real photon with $\omega=|{\bf q}|=q$ is absorbed by the nucleus, while in electron scattering a virtual photon is exchanged, where one can vary $\omega$ and $q$ independently.

In the cases of the photoabsorption and the electron-scattering process (see also Sections \ref{photon_results}, \ref{electron_results}), the cross section can be written in terms of a  so-called response function, which, in the inclusive unpolarized case, is defined as
\begin{equation} 
  R(\omega,q)=\int\!\!\!\!\!\!\!\sum_{\bar{0}f} \left|\langle \Psi_f|{\Theta}(q)|\Psi_0 \rangle \right|^2\delta\left(E_f-E_0-\omega \right).
\label{eq:rs}
\end{equation}
Here, $\Theta(q)$ is the electromagnetic operator, which can be directly one of the operators $(\rho, {\bf J})$  or can be just a multipole of them. $|\Psi_{0/f}\rangle$ are the ground state and the excited states of the Hamiltonian $H$, respectively. The symbol $\sum_{\bar{0}}$ indicates an average on the initial angular momentum projection, while
the symbol  $\int\!\!\!\!\!\sum_f$ corresponds to both a sum over discrete excited states and an integral over continuum eigenstates of the Hamiltonian. Indeed, $|\Psi_f\rangle$ may include not only bound excited states, but also states in the continuum where the nucleus is broken up into fragments. 

The calculation of continuum wave functions represents a challenging task especially in an inclusive process, where one  needs  information on all possible fragmentation channels of the nucleus at a given energy. 
To avoid the issue, one can use integral transforms, such as the Lorentz integral transform (LIT) technique~\cite{efros1994,efros2007}. Originally used in few-body calculations, the LIT technique is based on the calculation of the following integral of the response function $R(\omega,q)$,
\begin{equation}
    L(\sigma,\Gamma, q) = \frac{\Gamma}{\pi} \int d\omega\; \frac{R(\omega, q)}{(\omega-\sigma)^2 + \Gamma^2}\,,
\end{equation}
which can be shown to be the squared norm of the solution of a Schr\"odinger-like equation calculated using bound-state techniques. Once $L(\sigma,\Gamma)$ is calculated, a numerical inversion procedure allows one to recover $R(\omega,q)$, see Ref.~\cite{efros2007} for details. 

\vspace{1cm}
\subsection{Numerical solvers}
In order to calculate electromagnetic observables, we first need a numerical solution of the Schr\"odinger equation. 
In the applications discussed in Sections~\ref{photon_results}, \ref{sum_rules} and \ref{electron_results}, we will use either few-body or many-body solvers depending on the mass range $A$ of the addressed nuclei.

We obtain the bound-state and scattering-state wave functions for the $A=2$ problem by solving the partial-wave Lippmann-Schwinger equations for the Hamiltonian. The response functions are then calculated by directly evaluating the matrix elements of the electromagnetic operator in coordinate space.  

To calculate few-body problems with $2<A<8$ we use hyperspherical harmonics expansions.
In this framework, one expands the  $A$-body intrinsic wave function in terms of 
 hyperspherical harmonics ${\mathcal H}_{K}$ and hyperradial functions $R_{n}$ as
\begin{equation}
\Psi=\sum_{K }^{K_{max}}\sum_{n}^{n_{max}} \alpha_{n K}R_{n}(\rho_r){\mathcal H}_{K}(\Omega)\,,
\label{expans}
\end{equation}
where $\alpha_{n K}$ are the coefficients of the expansion and where for the sake of simplicity we omit  spin and isospin degrees of freedom. Here,  $\rho_r$ is the hyperradius while
 $\Omega$ is a set of hyperangles,  on which the hyperspherical harmonics ${\mathcal H}_{K}$ with grandangular momentum $K$ depend.
The expansion is performed up to a maximal value of hyperradial functions $n_{max}$ and a maximal value of grandangular momentum $K_{max}$. Reaching convergence in $n_{max}$ is typically not difficult. The expansion in hyperspherical harmonics is instead more delicate and one needs to ensure that the dependence of the calculated observables on this truncation is under control. To accelerate convergence, an effective interaction a la Lee-Suzuki can be introduced~\cite{EIHH}, obtaining the so-called effective interaction hyperspherical harmonics (EIHH) method, which allows to eventually achieve  sub-percentage accuracy in the $^4$He calculations of binding energies and electromagentic observables~\cite{LiMuli21}. Hyperspherical harmonics expansions can be conveniently used also to solve the  Schr\"odinger-like equation obtained when applying the LIT method described above. The interested reader can consult, e.g., Refs.~\cite{BN98,EIHH,efros2007,BaccaPastore2014,Ji_2018,LiMuli21} for more details.

For nuclei with $A\ge 8$ we use coupled-cluster theory. In this framework, for a given Hamiltonian $H$ one starts from a Slater determinant $|{\rm \Phi_0}\rangle$ of single particle states and assumes an exponential ansatz to construct the correlated
many-body wave function as 
\begin{equation}|\Psi_0\rangle=\exp{(T)}|{\rm \Phi_0}\rangle\,.
\label{anzatz}
\end{equation}
The operator $T$ is typically expanded in $n$-particle-$n$-hole excitations
(or clusters) as $T=T_1+T_2+\dots+T_A$.
Coupled-cluster theory is exact when the expansion of the $T$ operator
is considered up to A particles -- A holes excitations ($Ap$--$Ah$
) within  a model space determined by the number $N_{max}$ of oscillator shells  considered~\cite{hagen2009b}. Even though truncations are typically introduced, they can lead to a result  very  close to the exact one due to the exponential ansatz (\ref{anzatz}). Because the computational cost of this method scales polynomially with increasing mass number $A$, it is a very convenient solver for medium mass and even heavy nuclei~\cite{Hu:2021}.

For closed (sub-) shell nuclei, coupled-cluster theory truncated at the $2p$--$2h$ level,  in the so
called coupled-cluster singles and doubles (CCSD) scheme, captures about $90\%$ of the full correlation energy.  When
including  triples excitation, even at the leading order in the so-called CCSDT-1 scheme~\cite{watts1993},  one can obtain almost 99$\%$ of the correlation energy~\cite{bartlett2007,hagen2009b}. 
It has been shown that coupled-cluster theory can be also used in conjunction with the LIT method, where one can reduce the problem to the solution of a bound-state like equation of motion~\cite{bacca2013}.

\section{Uncertainty quantification}

In each of our computations of electromagnetic observables, the final accuracy will be controlled on the one hand by the
employed $\chi$EFT (determined by Hamiltonian and currents)  and on the other hand by the accuracy to which one can solve the few--body or many--body problem for a given Hamiltonian and current operator. 
Hence, in the following we will divide the sources of uncertainties in  two broad categories:
\begin{itemize}
    \item[($i$)] $\chi$EFT uncertainties;
    \item[($ii$)] Numerical uncertainties.
\end{itemize}

Among the uncertainties in ($i$), there are possible dependencies on the employed interaction or current model (including  cutoff dependencies), as well as uncertainties introduced by the truncation to a given order $\nu$ of the
employed $\chi$EFT, and uncertainties due to  extracting the LECs from experimental data or from lattice calculations.
If the LECs are well constrained by experimental data, the $\chi$EFT  uncertainty is typically dominated by the truncation error of the $\chi$EFT expansion. Regarding the latter, if the leading non-vanishing contribution to a calculated observable ${\mathcal O}$ enters at order $\nu_0$ 
and one is able to perform calculations that include all effects up to order $\nu_0+k$, one can naively expect to incur a relative error of  $\delta_{\CO}^{\chi\mathrm{EFT}}/\CO \approx (Q/\Lambda)^{k+1}$ from the neglected higher-order terms.
A more rigorous estimate can be obtained by using the calculated order-by-order results $\CO_\nu$ as ``data" to inform the uncertainty analysis. For example, using the simple algorithm proposed by Ref.~\cite{Epelbaum:2014sza}, the absolute truncation error can be estimated as 
\begin{equation}
\label{band}
 \delta_{\CO}^{\chi\mathrm{EFT}} = {\rm max}\left\{
\left(\frac{Q}{\Lambda}\right)^{k+1} \left|{\CO}_{\nu_0}\right|,
 \left(\frac{Q}{\Lambda}\right)^{k} \left|{\CO}_{\nu_0+1}-{\CO}_{\nu_0}\right|, 
 \ldots, \left(\frac{Q}{\Lambda}\right) \left|{\CO}_{\nu_0+k}-{\CO}_{\nu_0+k-1}\right|\right\}\,.
\end{equation}

More recently, Bayesian methods have been adopted for quantification of the $\chi$EFT truncation error~\cite{Schindler09,Furnstahl2015,Ekstrom2020,Epelbaum2020}. These methods start from Bayesian priors that encode naturalness~\footnote{As discussed in Ref.~\cite{Thomas}, care should be taken in attempts to quantify naturalness assumptions in Bayesian priors.} of the coefficients $\lbrace c_\nu\rbrace$ defined, using a suitable reference ${\CO}_\mathrm{ref}$, by  
\begin{equation}
\label{eq:eft_exp}
    {\CO}_{\nu_0+k} = {\CO}_\mathrm{ref}\sum_{\nu=0}^{k} c_\nu \, \left(\frac{Q}{\Lambda}\right)^\nu\,. 
\end{equation}
The priors are then updated using the calculated data to  arrive at a Bayesian posterior for the truncation error $\delta_{\CO}^{\chi\mathrm{EFT}}$. A key advantage of this approach is that the estimates  have a statistical interpretation, which allows us to validate the assumptions made and to easily combine truncation errors  with other sources of uncertainties such as fitting or random sampling of parameters. This opens a possible path for a complete and consistent accounting of theory uncertainty from all dynamical ingredients in the future. 

Regarding the uncertainties in ($ii$), the protocol to evaluate them will depend on the implemented numerical solver.
On the one hand, when performing a few-body calculations with hyperspherical harmonics, one needs to carefully take the convergence in $K_{max}$ into account. When using the LIT method, one also needs to consider the uncertainty of the inversion procedure.
On the other hand, when using coupled-cluster theory, one needs to account for at least two different patterns of convergence.
First, there is always a truncation on the model space controlled by the maximum number of harmonic oscillator (HO) shells $N_{max}$, which, in a sense, is analogous to the  $K_{max}$ in hyperspherical harmonics. If convergence in $N_{max}$ is reached, the results should in principle be independent of the underlying HO frequency $\hbar\Omega$ used for single particle states. However, in practice, one is always left with some residual $\hbar\Omega$ dependence which should be explored. Second, in coupled-cluster theory one has a cluster expansion of the operator $T$. Here, the most frequently adopted approximation is CCSD. When possible, one should include higher order excitations, such as leading order triples corrections with CCSDT-1. Finally, when using the LIT method, one incurs the extra numerical uncertainty coming from the inversion procedure.

In general, we expect  uncertainties of  ($ii$) to  be sub--percentage or at most one percent in light nuclei up to mass number 4, while  for medium--mass nuclei they may increase up to a few percent, depending  on the specific observable.  
Beyond the lightest nuclei, whether the uncertainties of ($i$) dominate over those of ($ii$) may, in principle, depend on the specific system/observable considered. Experience has shown so far that uncertainties related to the $\chi$EFT ($i$) are typically the largest. We will compare the specific contributions in each example below.


\section{Photoabsorption cross section}
\label{photon_results}

Photoabsorption cross sections have been extensively studied using ab-initio techniques, especially in the sector of light nuclei, see Ref.~\cite{BaccaPastore2014} and references therein. The photoabsorption cross section is related to the response function by 
\begin{equation}
\sigma_{\gamma}(\omega)= \frac{4\pi^2}{\omega} \alpha R_T(\omega,\omega)\,,
\label{cs_full}
\end{equation}
where $R_T(\omega,\omega)$ is the response function of Eq.~(\ref{eq:rs}) where the $\Theta$ operator is the  transverse  (with respect to photon propagation) part of the electromagnetic current operator ${\bf J}$ and where $\omega=q$. 
In the unretarded dipole approximation, the cross section can be obtained from 
\begin{equation}
\sigma_{\gamma}(\omega)=4\pi^2 \alpha \omega R_D(\omega)\,,
\label{cs}
\end{equation}
where $R_D(\omega)$ is the response function of the electric dipole operator ${\bf D}$ in the long wave length approximation. 

\vspace{2cm}

Below, we will discuss two examples. First, we will deal with  the radiative capture reaction $n ~ p \rightarrow \gamma {\mathrm d}$ reaction, which is important for astrophysics and is  related to the photoabsorption reaction $\gamma  {\mathrm d}  \rightarrow n p$ by time-reversal. Next, we will discuss the inclusive photoabsorption of $^4$He, for which we will present new original results obtained with chiral forces at four different orders, including an analysis of its uncertainties.

\vspace{0.5cm}
\subsection{The $n~p \leftrightarrow \gamma {\mathrm d}$ reaction}

\begin{figure}[hbt]
\centering
	\includegraphics[width=0.7\textwidth]{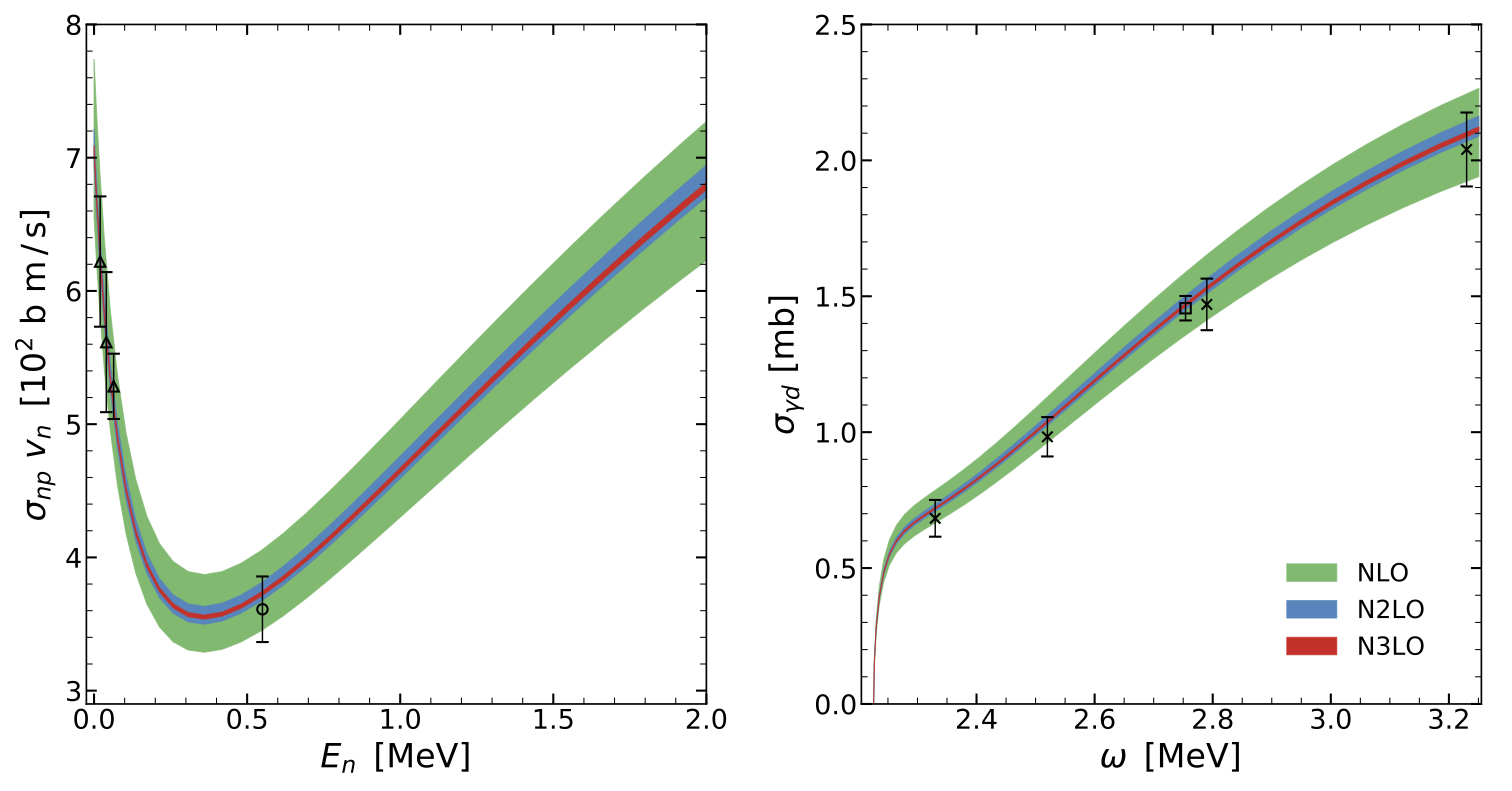}
	\caption{The product of $p(n,\gamma)d$ cross section $\sigma_{np}$ and the neutron speed $v_n$ versus the neutron energy $E_n$ (left panel); and the deuteron photodissociation cross section $\sigma_{\gamma d}$ as a function of the photon energy $\omega$ in the rest frame of the deuteron (right panel). The bands indicate 95\% Bayesian degree-of-belief intervals at the various orders. Experimental data are from Refs.~\cite{suzuki}~(triangles), \cite{nagai}~(circle), \cite{hara}~(crosses) and \cite{moreh}~(square). Experimental errors in beam-energy resolution are not shown.}
	\label{fig_npdg}
\end{figure}

The primordial Deuterium abundance, which is very well constrained by astronomical \cite{Cooke:2017cwo} and cosmological \cite{Planck:2018vyg} observations, can also be determined from nuclear physics by measuring or calculating the rates of the Deuterium production and burning reactions of the big-bang-nucleosynthesis network. While there is a reasonable  agreement between these at the moment \cite{Pisanti:2020efz}, a higher-precision comparison will search more rigorously for potential conflict which will be indicative of missing physics in one or the other and may even hint at new physics beyond the Standard Model. This elevates the importance of uncertainty quantification in the  
the primordial Deuterium production reaction, $n~p \rightarrow \gamma {\mathrm d}$.

In the relevant energy regime, $M1$ and $E1$ transitions are both important; we, therefore, evaluate the cross section using the full response function $R_T(\omega,\omega)$ with the one- and two-body current operators shown in Fig.~\ref{fig_H_J}. The uncertainties associated with the solution of the Schr\"odinger equation and other numerical approximations are negligible for this system. We therefore focus on $\chi$EFT uncertainties for this reaction. Working with fixed currents, we used the semi-local momentum-space-regularized chiral interactions of Ref.~\cite{Reinert:2017usi} to study the convergence properties of the $\chi$EFT expansion of the nuclear potential in Ref.~\cite{ACHARYA2022137011}. We employed the Gaussian Process (GP) error model developed in Ref.~\cite{melendez} to perform a Bayesian analysis of the $\chi$EFT convergence for observables that have parametric dependence on a kinematic variable, which in this case is the $np$ relative momentum. We performed detailed diagnostic checks to quantitatively assess the adequacy of the GP model and found that 
it described the observed convergence very well, which allowed us to extract reliable Bayesian posteriors for $\delta_{\CO}^{\chi\mathrm{EFT}}$ at various orders. 

In Fig.~\ref{fig_npdg}, we show the 95\% degree-of-belief bands for calculations at next-to-leading order (NLO), next-to-next-to-leading order (N2LO) and next-to-next-to-next-to-leading order (N3LO) obtained by using the leading order (LO) result as the reference $\CO_\mathrm{ref}$ (see Eq.~\eqref{eq:eft_exp}). We note that the theory uncertainty from the truncation of $\chi$EFT at N2LO and N3LO are much smaller than experimental errors at the energy range of astrophysical relevance. The uncertainty from truncation of the current operator is a subject of future study.

\vspace{0.5cm}
\subsection{The $\gamma ~^4${\rm He}$~ \rightarrow X$ reaction}
\label{sec:photo}

The photodisintegration cross section of $^4$He has been a focus of several past studies \cite{Ell96,Efros97,Barnea01,Qua04,Qua07,Doron06}. In this work, we provide new original results for this reaction obtained within the frameworks of $\chi$EFT using the EIHH~ \cite{EIHH,Barnea3NF,Barnea03} as a solver. We start from Eq.~(\ref{cs}) and keep the dipole operator fixed, while changing the nuclear interaction in the Hamiltonian implementing different orders in the chiral expansion. We work up to N2LO  with a maximally local version of the chiral interaction developed for the first time in Refs.~\cite{gez14,lyn16,lyn17}, which we previously adapted to the EIHH method in Ref.~\cite{LiMuli21}. 
In the same spirit of our work in the $n~p \leftrightarrow \gamma {\rm d}$ reaction, the uncertainty coming from the numerical solution of the Schrödinger equation is neglected here, since the EIHH method has been proven to be very precise for three- and four- body systems, with uncertainties that usually are below the percent level. 

\begin{figure}[hbt]
    \centering
	\includegraphics[width=0.95\textwidth]{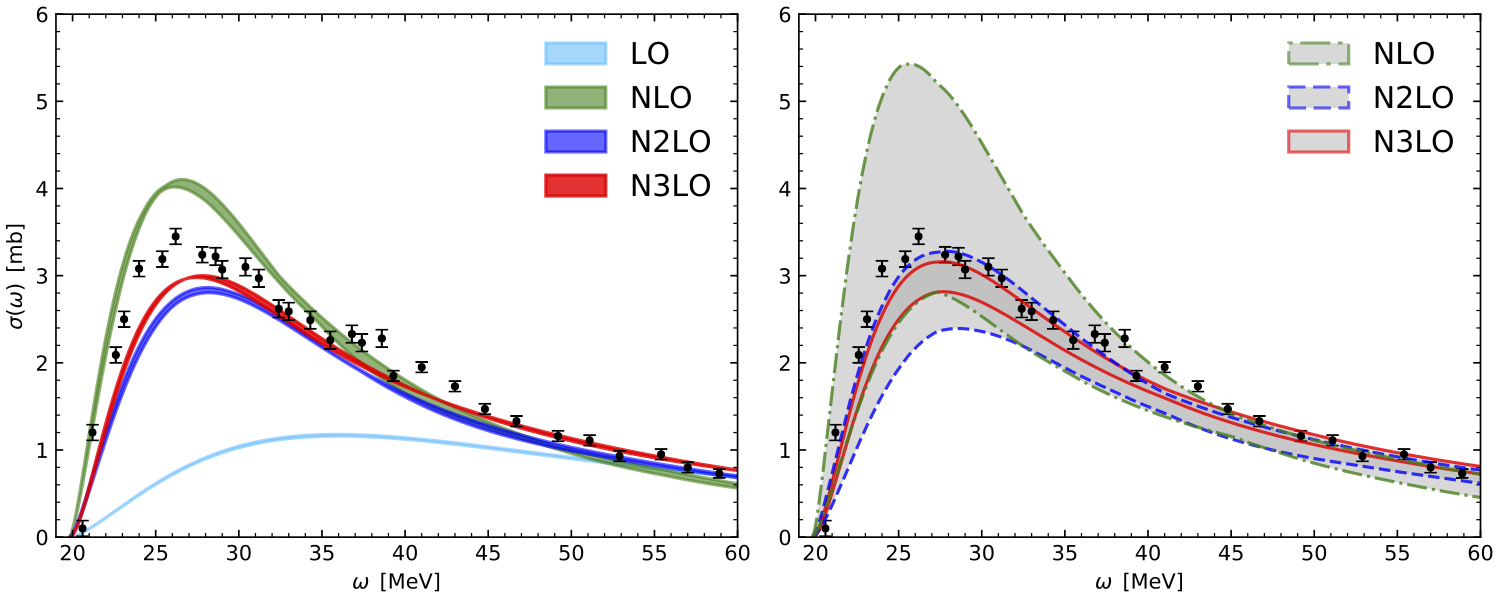}
	\caption{Inclusive $^4$He photoabsorption cross section calculated at different order in the chiral interaction. Left panel: bands display the numerical uncertainty in the inversion of the LIT. Right panel: bands display the $\chi$EFT truncation uncertainty, estimated using Eq.~(\ref{band}). The experimental data are taken from Ref.~\cite{Ark79}.} 
	\label{fig:4He}
\end{figure}

To bypass the explicit calculation of the continuum wave functions, the electric dipole response $R_D(\omega)$ of Eq.~(\ref{cs}) is obtained by first computing its LIT and then performing the inversion. This introduces a numerical uncertainty of the order of 1-2 $\%$, which can be seen in Figure~\ref{fig:4He} (left panel), where we present the calculation of $\sigma_\gamma(\omega)$ at LO, NLO, N2LO and N3LO in different colors. The width of the band is the uncertainty introduced by the inversion. 

To assess uncertainty coming from the truncation of the chiral expansion, we start from the calculations of $\sigma_\gamma(\omega)$ at the various chiral orders and implement the algorithm in Eq.~(\ref{band}), which  requires a choice for the expansion parameter $Q/\Lambda$. 
A reasonable choice for $Q$ is obtained by a smooth max-function (see Eq.~(46) of Ref~\cite{melendez}) of  $m_{\pi}$ and  $p_{\rm rel}=\sqrt{2\mu \omega}$, where $m_{\pi}$ is the pion-mass and $\mu$ is the reduced mass of the main photodisintegration channel, for which   we take $p - ^3$H.  For  the breakdown scale $\Lambda$ we take $500$ MeV. When implementing Eq.~(\ref{band}), we go beyond our calculations with local chiral interactions which go all the way up to N2LO, and also consider the partial N3LO calculation from Ref.~\cite{Qua07} and use it to estimate the uncertainty also at this order.

In Figure~\ref{fig:4He} (right panel), we show the cross section with corresponding $\chi$EFT uncertainty  at the NLO, N2LO and N3LO orders.  For every order the threshold energies are shifted to the experimental value.
Clearly, the $\chi$EFT errors account for the largest portion of the overall uncertainty budget with respect to the numerical inversion uncertainty, which are therefore not even included in the  right panel of Fig.~\ref{fig:4He}. The $\chi$EFT truncation errors are such that the calculated photoabsorption cross section at each order is consistent with the previous order within its uncertainties, as well as with the experimental data from Refs.~\cite{Ark79}.
 At NLO we get an  uncertainty at the cross section maximum of roughly 30\% (half width), while at  N2LO it is 15\% (half width). Finally, the N3LO band, which is roughly 5\% (half width), is located slightly below the shown experimental data. To facilitate comparison of theory with experiment, we have chosen to show only one representative set of data \cite{Ark79}, which covers a wide range in energy. More data exist than are shown here, see, e.g., Ref.~\cite{BaccaPastore2014} and references therein.

\vspace{0.5cm}
\section{Electromagnetic sum rules }
\label{sum_rules}

Starting from the nuclear response function, one can compute electromagnetic sum rules, i.e., the moments of the response function of Eq.~(\ref{eq:rs}) interpreted as a distribution function. These quantities are defined as 
\begin{equation}
    m_n (q)= \int d\omega\;\omega^n R(\omega, q), 
    \label{mn_def}
\end{equation}
where $n$ is an integer. Sum rules can be calculated directly from the LIT. Since for $\Gamma \rightarrow 0$, the limit of a Lorentzian corresponds to a delta function, we get
\begin{equation}
   L(q, \sigma, \Gamma\rightarrow 0) = \int d\omega\;R(\omega,q)\delta(\omega -\sigma) = R(q,\sigma)\,.
\end{equation}
This means that the moments of $R(\omega,q)$ can be obtained from the following expression
\begin{equation}
    m_n (q) = \int d\sigma\; \sigma^n L(q, \sigma, \Gamma\rightarrow 0).
    \label{limit}
\end{equation}
As illustrated in Ref.~\cite{miorelli2016}, this procedure is equivalent to the computation and subsequent integration of the response. Moreover, this strategy does not require  an inversion, which represents an additional source of uncertainty.

Among the sum rules, the electric dipole polarizability $\alpha_D$ is an interesting one, as it is correlated to parameters in the neutron-matter equation of state~\cite{roca-maza2013}. The electric dipole polarizability can be obtained starting from the inverse-energy weighted sum rule
\begin{equation}
    \alpha_D = 2\alpha \int d\omega\; \frac{R_D(\omega)}{\omega} = 2\alpha m_{-1}\,,
\label{alphaD_m-1}
\end{equation}
where $m_{-1}$ is calculated using Eq.~(\ref{mn_def}) are $R=R_D(\omega)$ is the  response function of the dipole operator in the long-wave length approximation. From Eq.~(\ref{alphaD_m-1}), it is clear that the polarizability is dominated by the low-energy part of the response function. 

\begin{figure}[hbt]
    \centering
	\includegraphics[width=0.7\textwidth]{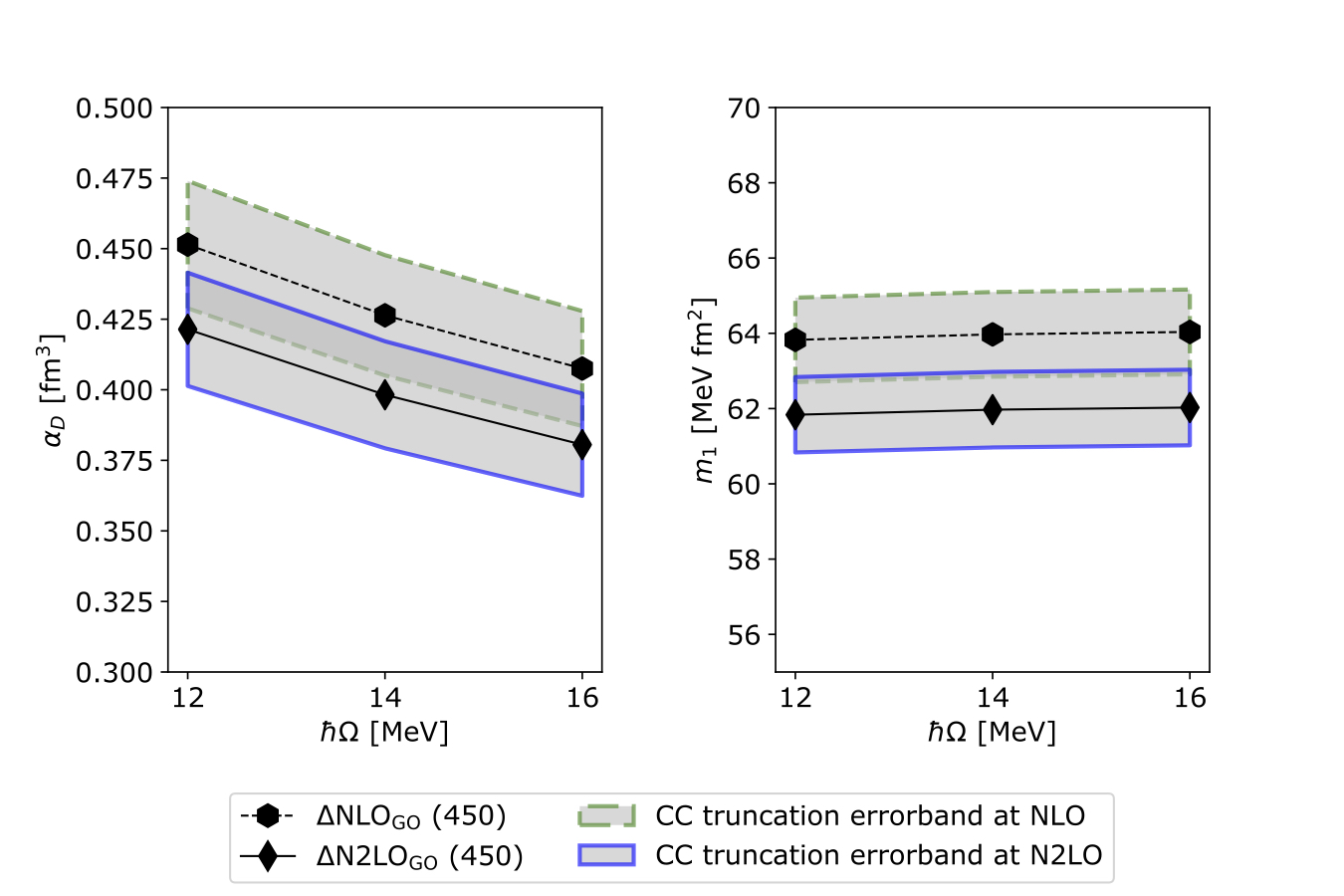}

	\caption{The $\hbar\Omega$-convergence pattern of $\alpha_D$ and $m_1$ for $^8$He calculated with $\Delta$NLO$_{\mathrm{GO}}(450)$ and $\Delta$N2LO$_{\mathrm{GO}}(450)$ at fixed $N_{max} = 14$. The green and blue bands indicate the CC truncation uncertainty. The black points are the results obtained including 3$p$-3$h$ excitations in both the ground- and excited-state computations. }
	\label{fig:8He}
\end{figure}

In a recent work~\cite{bonaiti2022}, we performed coupled-cluster computations of dipole-excited state properties of the halo nucleus $^8$He, focusing on $\alpha_D$ and the energy-weighted sum rule $m_1$ using $\chi$EFT potentials derived at N2LO. Our calculations included an estimate of the theoretical uncertainty related to the model space convergence in $N_{max}$ and to the truncation of the coupled-cluster expansion, according to the strategy illustrated in Ref.~\cite{simonis2019}.
Regarding the first source of uncertainty, the maximum available model space is  $N_{max} = 14$, so we consider
the residual $\hbar\Omega$-dependence at this $N_{max}$ as the uncertainty in the model space expansion.
To assess the uncertainty in the coupled-cluster expansion, we take two different approximation schemes, the CCSD and the CCSDT-1, since we have no higher order coupled-cluster approximations available. The truncation uncertainty is then estimated taking half of the difference between the CCSD and CCSDT-1 results. 
The two contributions are then summed in quadrature. 

To complement our previous analysis, we consider in this work the dependence on the order of the $\chi$EFT  expansion in the case of the $\Delta$-full interaction model, by providing a new calculation at a lower order (NLO).
In Figure \ref{fig:8He}, we show the $\hbar\Omega$ convergence pattern of $\alpha_D$ and $m_1$ for the $\Delta$NLO$_{\mathrm{GO}}(450)$ and $\Delta$N2LO$_{\mathrm{GO}}(450)$ potentials~\cite{jiang2020}, indicating with bands the contribution of the coupled-cluster truncation uncertainty. In the case of the dipole polarizability, the theoretical error receives substantial contributions from both the many-body method and the residual dependence on the coupled-cluster convergence parameters. The polarizability is sensitive to the outer part of the nuclear wave function, and this makes the convergence slower for a loosely-bound system like $^8$He. $\Delta$NLO$_{\mathrm{GO}}(450)$ predicts a slightly larger polarizability with respect to $\Delta$N2LO$_{\mathrm{GO}}(450)$. Taking into account the uncertainty budget coming from the many-body solver (around $7\%$ of the central value), the two results come out to agree within errobars. 

The situation changes when turning to the energy-weighted sum rule. Here the overall uncertainty is dominated by the coupled-cluster truncation and it is estimated to be below $2\%$. Also in this case $\Delta$NLO$_{\mathrm{GO}}(450)$ leads to a larger value for $m_1$. However, due to the smooth convergence of this observable, the difference between the two chiral orders, amounting to $3\%$, can be better appreciated than in the case of the polarizability. At the moment it is possible only two compute two orders in the $\chi$EFT expansion, namely the NLO and N2LO, therefore we refrain from using the algorithm of Eq.~(\ref{band}) in this case. Clearly, the uncertainty analysis is then less sophisticated than for the  $A=2,4$ nuclei, but it is reassuring to see that the  NLO and N2LO error bands overlap.


\section{Electron scattering cross section}
\label{electron_results}

\begin{figure}[hbt]
	\includegraphics[width=0.5\textwidth]{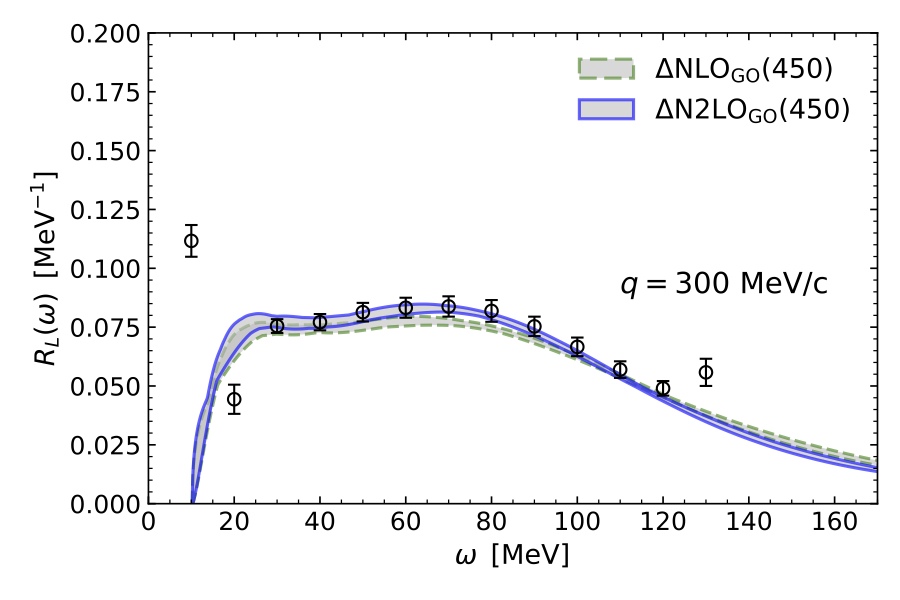}
	\includegraphics[width=0.5\textwidth]{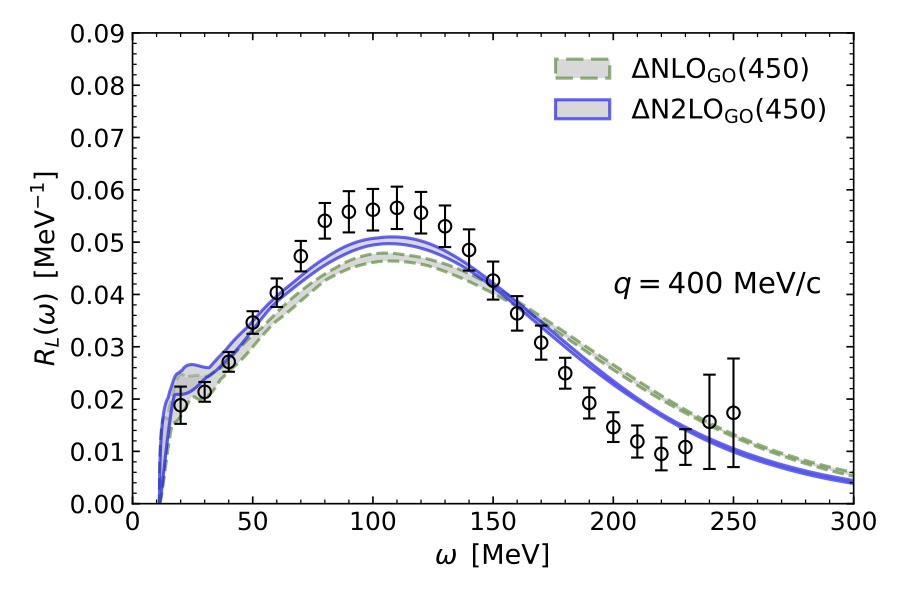}
	
	\caption{Longitudinal response functions of $^{40}$Ca for the momentum transfer $q=300$ MeV/c (left panel) and $q=400$~MeV/c (right panel). Two orders of the chiral expansion of the nuclear Hamiltonian are shown. The uncertainty band originates from the inversion procedure of the LITs. The experimental data are taken from Ref.~\cite{Williamson:1997zz}.}
	\label{fig:40Ca_EFT}
\end{figure}
Electron scattering has proven to be a powerful tool to investigate the nuclear structure and dynamics at various energy scales and for different systems. 
Very recently we started investigating the region of the quasielastic peak which becomes a dominating mechanism for the momentum transfer of the order of hundreds of MeV, below the pion production threshold.
The inclusive electron-nucleus cross section can be expressed as
\begin{equation}
    \frac{d^2\sigma}{d\Omega d\omega} = \sigma_{\mathrm{Mott}} \left( \frac{{\bf q}^2-\omega^2}{{\bf q}^2}R_L(\omega,q) + \left(\frac{{\bf q}^2-\omega^2}{2{\bf q}^2} + \tan^2\frac{\theta}{2}\right) R_T(\omega, q) \right)
\end{equation}
with the longitudinal and transverse response functions $R_{L/T}$ and the scattering angle $\theta$. The response functions can be disentangled experimentally via the Rosenbluth separation technique. From the theoretical point of view, it is convenient to investigate first the longitudinal component,  which is the response function of Eq.~(\ref{eq:rs}) where the operator $\Theta(q)$ is the charge operator
\begin{equation}
    \rho(q) = \sum_{i=1}^Z e^{iq(r_i-R_{cm})}\, .
\end{equation}
The operator structure of $\rho$ is simpler than that of the electromagnetic current ${\bf J}$ and two-body contributions  appear at a high order in the chiral expansion (see Fig.~\ref{fig_H_J}), so that it can be  neglected if performing studies up to N2LO.
While the ab-initio calculations of $R_L$ in light systems were performed in several theoretical frameworks, we recently extended these studies to the region of medium-mass nuclei~\cite{Sobczyk:2021dwm}. We focused on $^{40}$Ca, for which Rosenbluth separated response functions are available, using two different N2LO potentials~\cite{ekstrom2015a,jiang2020}. Here, we complement our uncertainty analysis by performing a new calculation with an NLO potential.

Similarly to the photoabsorption considered in Sec.~\ref{sec:photo}, the calculation of $R_L$ requires computing the LITs which afterwards have to be inverted, introducing an additional source of uncertainty with respect to the sum-rule calculation. We obtain the LITs using coupled-cluster theory within the CCSD approximation. The role played by $3p-3h$ excitations will be a topic of the future investigation.
We calculate $R_L$ using a single model space of $N_{max}=14$ and harmonic oscillator frequency $\hbar\Omega=16$ MeV. In our previous work~\cite{Sobczyk:2021dwm} we varied the frequency of the underlying harmonic oscillator basis and its size and we found that the LITs are already well converged. In this situation, the numerical uncertainty is driven by the inversion procedure which is represented by the band shown in Fig.~\ref{fig:40Ca_EFT}.

To assess the uncertainty coming from the $\chi$EFT expansion we look at the dependence on the order of expansion of the $\Delta$-full potential~\cite{jiang2020} at NLO and N2LO. 
In Fig.~\ref{fig:40Ca_EFT} we present $R_L$ for $q=300$ MeV/c (left panel) and $q=400$ MeV/c (right panel). At $q=300$ MeV/c the predictions of $\Delta$NLO$_{\mathrm{GO}}(450)$ and $\Delta$N2LO$_{\mathrm{GO}}(450)$ agree to great extent  within the uncertainty bands and  with the data. In contrast, at  $q=400$ MeV, where the uncertainty bands of the $\Delta$NLO$_{\mathrm{GO}}(450)$ and $\Delta$N2LO$_{\mathrm{GO}}(450)$ overlap less and where the agreement with data slightly deteriorates. When comparing the two interactions, we see that the $\Delta$N2LO$_{\mathrm{GO}}(450)$ leads to a slightly higher and narrower quasielastic peak with respect to the $\Delta$NLO$_{\mathrm{GO}}(450)$ (the difference of around $8\%$ in the peak for $q=400$ MeV/c),  bringing the results closer to the data as the chiral order increases.
Because a quantitative analysis would require more than two orders of the $\chi$EFT Hamiltonian, we refrain here from applying Eq.~(\ref{band}), which would only contain one term.

At the qualitative level,
we observe that the size of the uncertainties of kind $(i)$ and $(ii)$ are comparable, and those of kind $(i)$ seem to depend on the momentum transfer and grow at larger $q$ value. This is, after all, not surprising, because $\chi$EFT is expected to work better at low momenta than at higher momenta.




\section{Conclusions}

In this paper, we review the recent progress made in  uncertainty quantification for ab-initio calculations of electromagnetic observables focusing on the one hand on our recent results and on the other hand providing also new original results to complement the uncertainty analysis. We show several examples where nuclei of different masses are scrutinized.

We first  showcase the recent computations of the $n~p \leftrightarrow \gamma {\mathrm d}$ reaction, where an uncertainty analysis of the $\chi$EFT truncation with Bayesian tools was implemented. Then, we show new results for the photoabsoprtion cross section of $^4$He computed with $\chi$EFT potentials at LO, NLO and N2LO. The uncertainty quantification we present is based on the use of Eq.~(\ref{band}) and pushed to N3LO using the results from Ref.~\cite{Qua07}.
For both these examples in the sector of light nuclei, we find that numerical uncertainties are negligible and the bulk of the error stems from the truncation of the $\chi$EFT expansion. 
Next, we discuss sum rules in the exotic $^8$He nucleus, where we confront the existing calculation at N2LO with a new computation at NLO in the $\chi$EFT expansion using $\Delta$ degrees of freedom. Here, we see that numerical uncertainties and $\chi$EFT truncation errors are comparable in size. Finally, we show results for the longitudinal response function of $^{40}$Ca using the same interactions we used for $^8$He.
Also in this case, the uncertainty stemming from the $\chi$EFT truncation seems comparable to that coming from the numerical solver. It is important to note here that we are not yet able to fully account for the numerical uncertainties, because we have not yet included $3p$--$3h$ excitations. Furthermore, we only have two orders in the $\chi$EFT so a quantitative uncertainty cannot yet be reliably estimated. Interestingly, we qualitatively observe a momentum-transfer dependence in the difference between the calculation at NLO and N2LO, which is not unexpected given that $\chi$EFT is a low-momentum expansion. A precise quantitative description of the dependence of the $\chi$EFT expansion on the momentum transfer, which is obscured by the fact that we use phenomenological form factors to represent photon-nucleon vertices, is a subject of future study. 

Clearly, the level of sophistication of our uncertainty quantification is higher for lighter nuclei and decreases as the mass grows. The most rigorous analysis was performed for $A=2$, where we were able to express the truncation errors as Bayesian degree-of-belief intervals. For the range of $A\simeq 4$ one can expect that a Bayesian analysis will be implemented in the future. A quantitative analysis of nuclei with $A\ge 8$ will need more effort. We expect LO calculations to be far from experimental data for these nuclei, but if one wants to go beyond N2LO in the $\chi$EFT expansion, one would need consistent potentials that are soft enough for many-body calculations to converge. Moreover, to fully assess uncertainties in electromagnetic observables, one must also consider the $\chi$EFT expansion in the current and in the interaction simultaneously. Finally, in the future statistical approaches for the variation of the LECs such as those shown in Ref.~\cite{Hu:2021} should be applied  broadly to the study of electroweak dynamical observables, such as response functions and cross sections.

\section*{Conflict of Interest Statement}

The authors declare that the research was conducted in the absence of any commercial or financial relationships that could be construed as a potential conflict of interest.

\section*{Author Contributions}

All authors contributed in equal parts to this paper. BA  led in the deuteron calculations, SSL  the $^4$He calculations, FB the $^8$He calculations, and JES together with BA performed the $^{40}$Ca calculations.
While SB took the main responsibility for the drafting of the paper, all authors contributed to the writing of the manuscript.

\section*{Funding}
This work was supported by the
  Deutsche Forschungsgemeinschaft (DFG) through Project-ID 279384907 - SFB 1245 and through the Cluster of Excellence ``Precision Physics,
  Fundamental Interactions, and Structure of Matter" (PRISMA$^+$ EXC
  2118/1) funded by the DFG within the German Excellence Strategy
  (Project ID 39083149). BA's work at ORNL is supported by the Neutrino Theory Network Fellowship Program (Grant No. DE-AC02-07CH11359). 
  JES acknowledges the support of the Humboldt Foundation through a Humboldt Research Fellowship for Postdoctoral Researchers and funding from the European Union's Horizon 2020 research and innovation programme under the Marie Skłodowska-Curie grant agreement No. 101026014.
  Computer time was provided by the Innovative and Novel Computational Impact on Theory and Experiment (INCITE) program and by the supercomputer Mogon at Johannes Gutenberg Universit\"at Mainz. This research used resources of the Oak Ridge Leadership Computing Facility located at ORNL, which is supported by the Office of Science of the Department of Energy under Contract No. DE-AC05-00OR22725.

\section*{Acknowledgments}
We would like to acknowledge  Weiguang Jiang for sharing the matrix elements for the  $\Delta$NLO$_{\mathrm{GO}}(450)$ potential with us.  We would like to thank Nir Barnea and Gaute Hagen for access to the hyperspherical harmonics and coupled-cluster codes, respectively. Finally, we would like to thank Thomas R. Richardson for a critical reading of the manuscript. 

\bibliographystyle{frontiersinHLTH&FPHY} 
\bibliography{master}

\end{document}